# Thermal Stability and Structural Relaxation Kinetics of Commercially Pure Titanium

**Y. Semerenko [1], N. Galtsov [1], D. Hurova [1], T. Bednarchuk [2], P. Zinoviev [1], V. Zoryansky [1], V. Moskalenko [1], R. Smolianets [1], A. Smirnov [1], Y. Pohribna [1], N. Aksenova [3]**

*[1] B.Verkin Institute for Low Temperature Physics and Engineering of NAS of Ukraine, 47 Nauky Ave., Kharkiv, 61103, Ukraine*
*[2] Institute for Low Temperatures and Structure Research, Polish Academy of Sciences, P.O. Box 1410, 50-950 Wroclaw, Poland*
*[3] V.N. Karazin Kharkiv National University, 4 Svobody Sq., 61022 Kharkiv, Ukraine*

This work explores the structural relaxation of cryorolled commercially pure titanium subjected to subsequent annealing. Nanostructured specimens with an average grain size of ~50 nm were produced by plastic deformation via cryogenic rolling at 100 K to a true strain of 1.20, followed by annealing at 425 K, 525 K, 725 K, and 945 K. Microstructural evolution was examined using electron microscopy and X-ray diffraction.

Cryogenic rolling was found to induce pronounced grain fragmentation, with coherent scattering regions (CSRs) of 30-50 nm dominating the substructure. X-ray diffraction analysis revealed that annealing drives crystallite growth: a slight increase at 525 K, accelerated coarsening at 725 K, and complete recrystallization at 945 K. The kinetics of grain growth follow the classical Lifshitz–Slyozov–Wagner model, with a growth exponent of ~ 4, reflecting the role of triple junctions and defect-controlled boundary diffusion. The activation energy, estimated at ~97 kJ/mol, corresponds to typical values for grain boundary diffusion in α-Ti, confirming that cryogenic deformation produces a highly non-equilibrium microstructure whose relaxation is strongly temperature dependent.

## 1. Introduction

A well-established route to enhancing titanium's strength is grain refinement - reducing the grain size down to the nanoscale (see Fig. 1). This approach is widely recognized in materials science for its ability to raise hardness and yield strength through mechanisms associated with grain boundary strengthening [1-3].

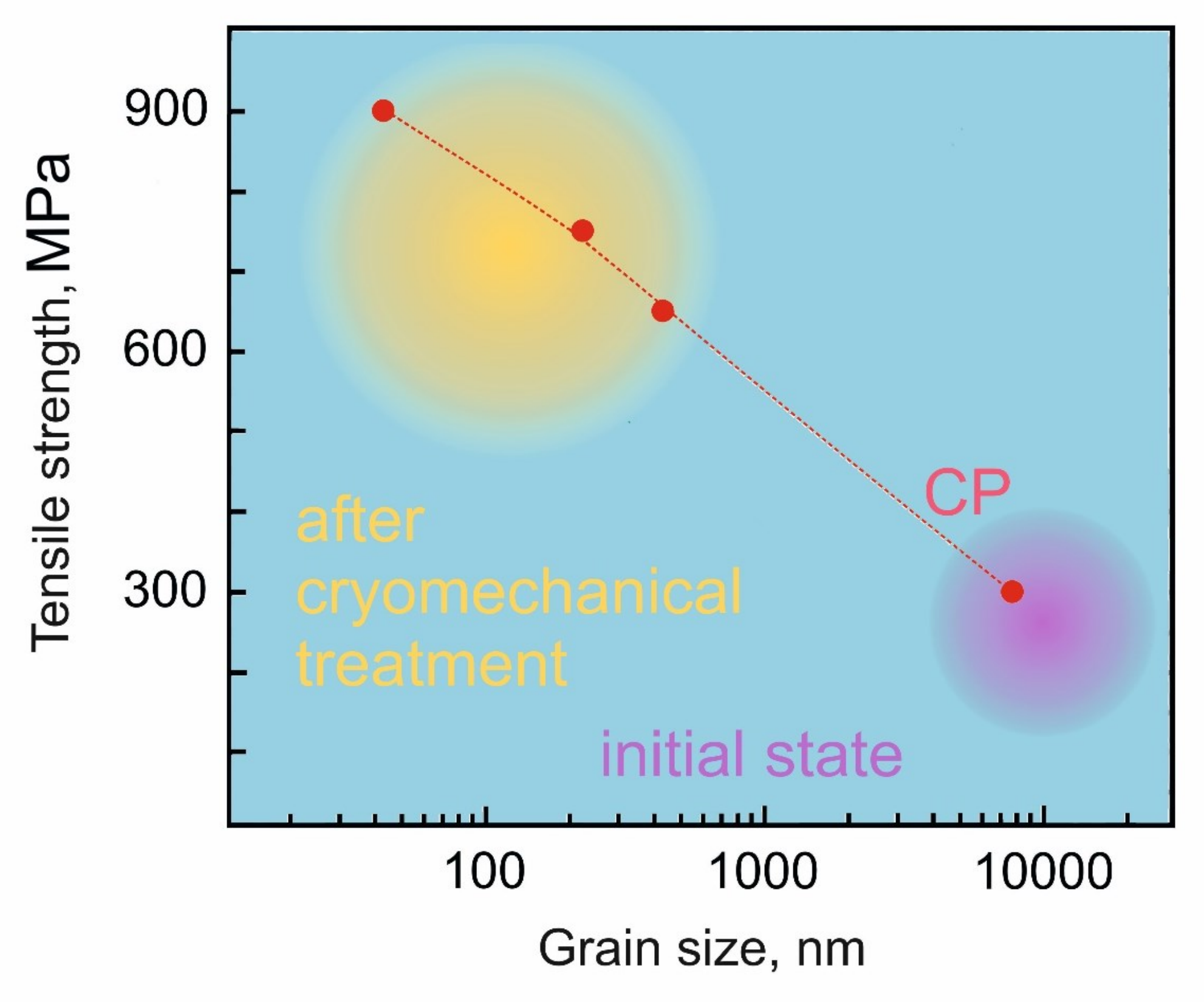

**Fig. 1.** The dependence of tensile yield strength on grain size for Commercially Pure (CP) titanium VT1-0 (analogous to ASTM Grade 2) at room temperature. In the nanostructured state, CP titanium exhibits a tensile strength reaching up to 900 MPa. These values are approximately three to five times greater than those observed in coarse-grained titanium. The figure provides a schematic overview compiled from published data sources [4-8]

Among the various processing techniques available, severe plastic deformation stands out as one of the most significant and widely applied methods. This approach enables the production of bulk materials with an ultrafine-grained structure (average grain size exceeding 200 nm) [9].

On the other hand, cryomechanical processing represents a highly promising route for achieving a nanocrystalline state in titanium. This method involves subjecting the material to plastic deformation at cryogenic temperatures, which promotes the development of ultrafine grains and enhances mechanical performance while preserving ductility [10, 11]. The cryomechanical approach is gaining increasing recognition due to its use of standard metal-forming techniques, as well as its scalability and ease of integration into existing manufacturing processes [4].

One of the most effective ways to fine-tune and stabilize the balance between strength and ductility in titanium subjected to cryomechanical treatment is through carefully controlled post-deformation annealing [12, 13]. This thermal post-treatment allows internal stresses to relax and promotes either partial recovery or full recrystallization, depending on the specific parameters applied. By carefully tuning the annealing temperature and duration, the microstructural features - such as grain size and dislocation density - can be adjusted to optimize the mechanical performance of the material for targeted applications. Controlled annealing therefore plays a pivotal role in achieving the desired balance of high strength and sufficient plasticity in nanostructured titanium.

This study focuses on the effects of controlled annealing on the evolution of structural properties in nanocrystalline (NC) specimens with grain size 30–50 nm of commercially pure titanium VT1-0 (containing 0.28% impurities, including 0.06 wt% Fe, 0.12 wt% O, 0.02 wt% N, 0.01 wt% C, and 0.003 wt% H; equivalent to ASTM Grade 2). The NC samples were produced via cryomechanical treatment. While their mechanical properties have been thoroughly investigated in previous studies [4, 14-19], their structural properties has remained less explored.

The aim of this research is to evaluate how different post deformation thermal exposures influence microstructural stability and dislocation configurations characteristics. Particular emphasis is placed on the kinetics of structural relaxation and defect dynamics during thermal and temporal evolution.

## 2. Experimental Methods

### *2.1 Cryogenic Temperature Rolling*

The substructural state of the specimens was generated through plastic deformation by cryogenic rolling (see Fig. 2) at approximately 100 K, a process referred to as Cryogenic Temperature Rolling (CTR). Rolling was performed to achieve a true strain of about $|e| = 1.2$. The true strain after rolling was determined using the formula $|e| = \ln\left( t/t_0 \right)$, where $t$ and $t_0$ denote the final and initial thicknesses of the rolled plate, respectively. After deformation, the specimens underwent annealing at 425 K, 525 K, 725 K, and 945 K to study the processes of microstructural evolution and relaxation.

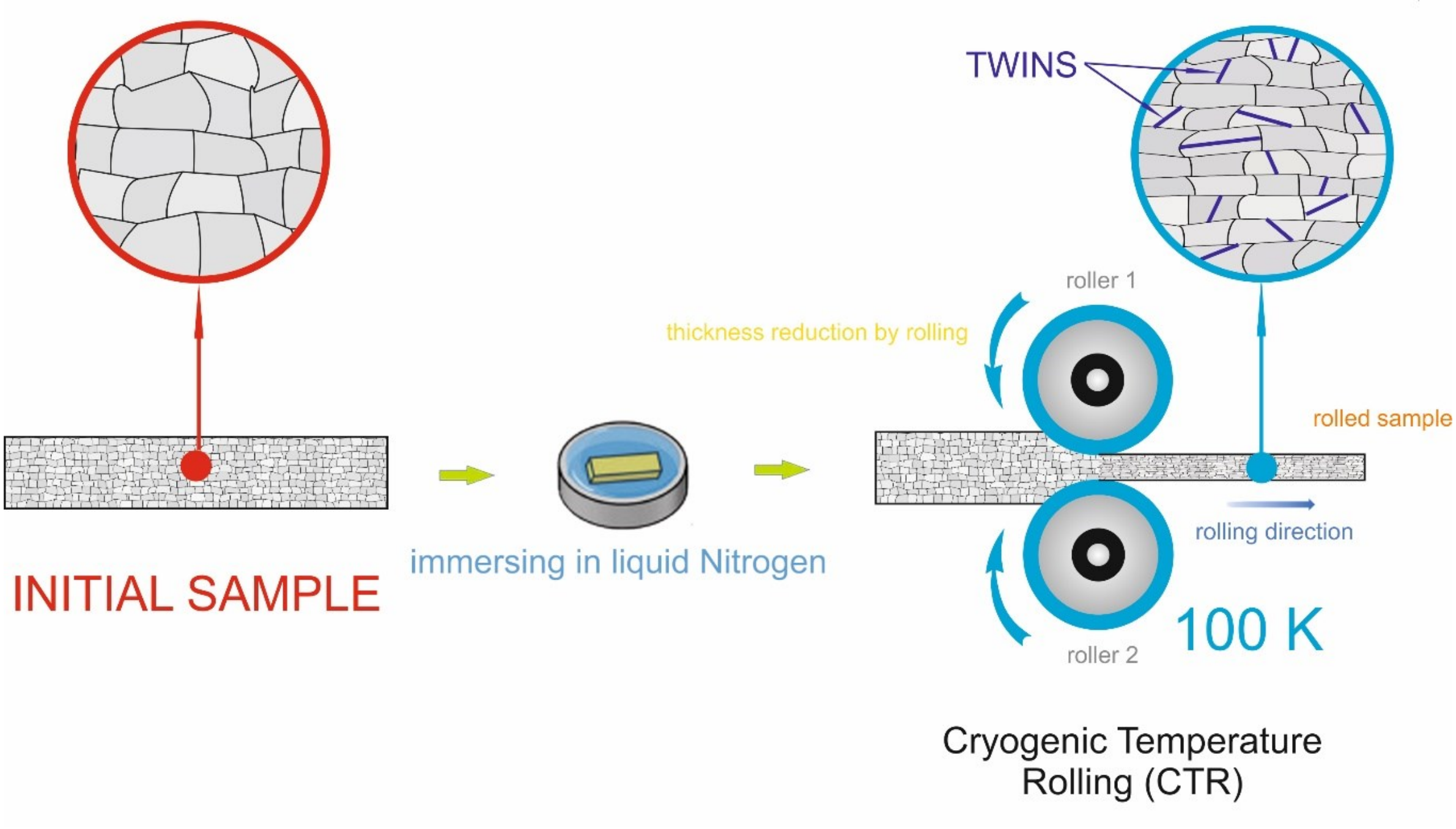

**Fig. 2.** Diagram of the rolling-induced deformation state formation in VT1-0 titanium specimens

The concept of cryomechanical treatment builds upon the findings reported in [4, 20, 21], which explored the fundamental physical mechanisms underlying plastic deformation of titanium at cryogenic temperatures. Under ambient conditions, plastic deformation in titanium - characterized by its Hexagonal Close-Packed (HCP) crystal structure - occurs primarily through prismatic slip along $\{10\bar{1}0\}\langle 11\bar{2}0\rangle$, with additional contributions from basal $(0001)\langle 11\bar{2}0\rangle$ and pyramidal $(0001)\langle 11\bar{2}0\rangle$ slip systems, as well as mechanical twinning, most commonly of the $\{10\bar{1}2\}$ and $\{11\bar{2}2\}$ types

At low temperatures, mechanical twinning becomes a dominant complementary mechanism of plastic deformation. In fact, nearly all grains within a polycrystalline aggregate undergo twinning even at relatively small strains. This behavior is promoted by the wide range of twinning systems that are activated under cryogenic conditions, together with the increased tendency for twinning itself. Both extension twins (e.g., $\{11\bar{2}2\}$ and $\{10\bar{1}2\}$), which elongate grains along the *c*-axis, and compression twins (e.g., $\{11\bar{2}2\}$ and $\{11\bar{2}4\}$), which shorten grains along the *c*-axis, are observed

The simultaneous activation of multiple twinning systems results in pronounced grain refinement. Further grain fragmentation arises from the extensive formation of secondary and tertiary twins, the intersection of twin boundaries, twinning within accommodation zones at the

twin–matrix interface, and interactions between twins and grain boundaries. Together, these processes generate a highly refined microstructure, highlighting the unique effectiveness of cryomechanical treatment in producing nanostructured states in titanium.

A more detailed description and justification of the cryorolling method applied to titanium is provided in [4]. That work offers a comprehensive discussion of the principles of low-temperature plastic deformation, with particular emphasis on mechanisms of grain refinement, defect generation, and stabilization of nanostructured states. The methodology is supported both theoretically and experimentally, demonstrating that cryogenic rolling not only enhances the strength of titanium but also preserves its ductility by suppressing dynamic recovery processes.

### *2.2 Transmission Electron Microscopy*

Structural examination was conducted using Transmission Electron Microscopy (TEM). Samples were prepared by spark cutting and subsequently polished to a thickness of about 0.25 mm. Final thinning to perforation was achieved in an electrolyte composed of perchloric acid mixed with butyl and methyl alcohols in a 1:6:10 volume ratio, cooled to 233 K. Observations were performed on the electron-transparent regions surrounding the perforations.

The deformation microstructure was analyzed with an EM200 transmission electron microscope operating at 175 kV. Dark-field images were obtained by selecting portions of three closely spaced diffraction rings $\{10\bar{1}0\}, (0002)$ and $\{10\bar{1}1\}$ - in the micro-diffraction pattern using an aperture diaphragm. Coherent Scattering Regions (CSRs) visible in the dark-field negatives were classified according to size, and a histogram of their distribution was constructed based on 800-1300 individual measurements.

Electron microscopy revealed that the substructure of cryogenically deformed titanium is predominantly composed of Coherent Scattering Regions (CSRs) with characteristic dimensions of about 30–50 nm (Fig. 3b). Following cryogenic deformation to a true strain of $|e| = 1.2$, CSRs are observed mainly in clustered formations. With further increase in strain, these clusters progressively break apart, leading to a more homogeneous distribution of CSRs throughout the microstructure.

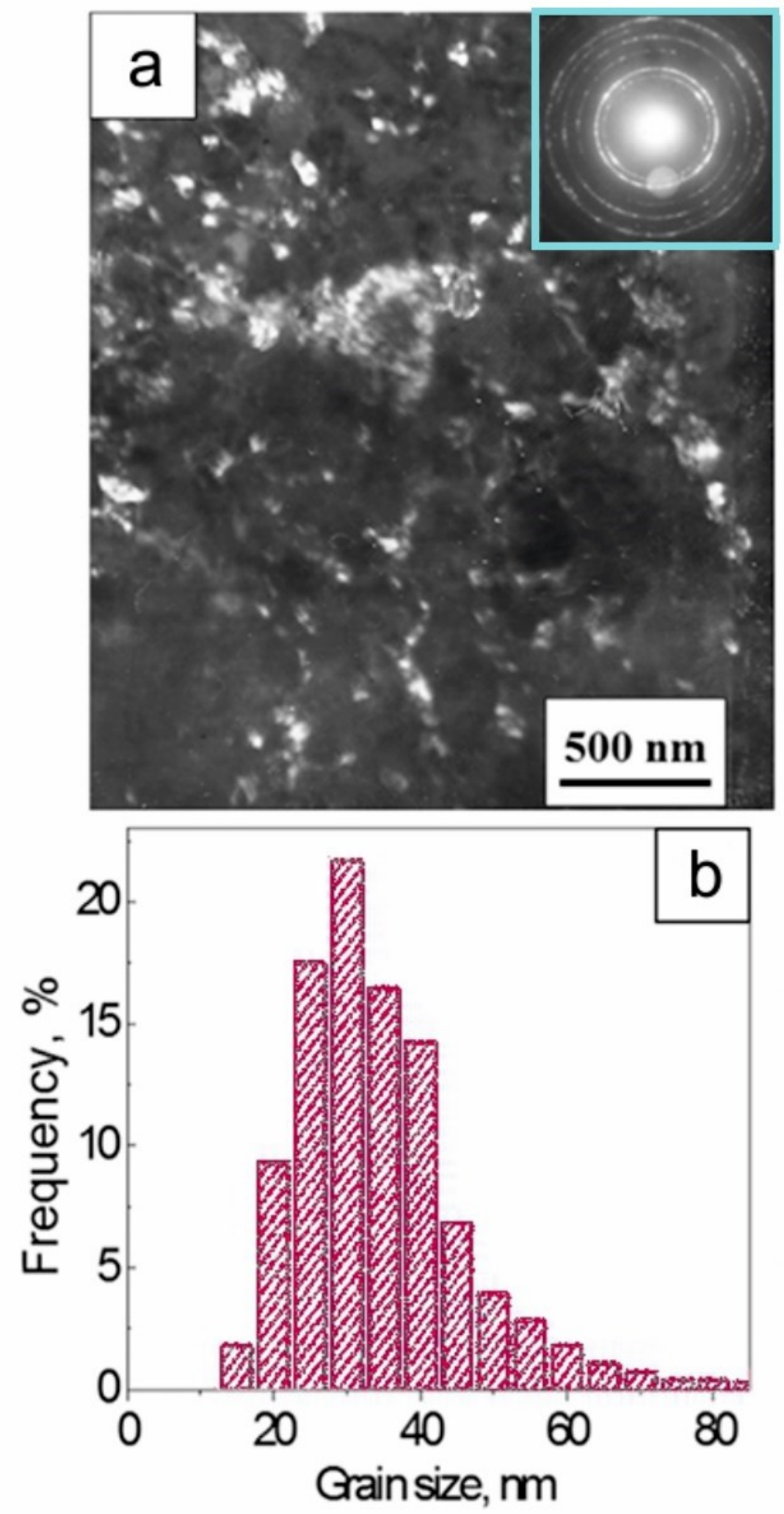

**Fig. 3.** TEM dark-field micrograph (a) and histograms of the grain size distribution (b) for samples after cryorolling to a strain $|e| \sim 1.2$ (state **1**). The inset in Fig. 3a displays the selected area electron diffraction (SAED) pattern. The presence of distinct rings confirms the polycrystalline nature of the material with fine-grained structure.

The formation of CSRs is attributed to repeated twinning processes occurring in titanium under low-temperature deformation [21]. This mechanism provides a consistent explanation for the emergence and evolution of nanoscale structural features in cryogenically rolled titanium. The nanocrystalline structure of commercially pure titanium VT1-0 produced by cryogenic deformation has been extensively studied in earlier works [4, 21, 22] using bright-field and dark-field TEM as well as X-ray diffraction. These complementary techniques offered detailed insight into microstructural characteristics such as grain refinement, defect distribution, and CSR development, thereby establishing a comprehensive understanding of the nanoscale architecture of cryogenically processed titanium.

## *2.3 X-ray diffraction analysis*

X-ray structural studies of the samples were performed using a DRON-2 serial diffractometer (NPO Bourevestnik, Saint Petersburg, USSR) with copper anode radiation and a nickel filter at room temperature. Measurements were carried out according to the scanning scheme $\theta - 2\theta$ with collimating slits [15]. The scattering intensity was recorded continuously over the specified angular range $4^\circ \leq 2\theta \leq 140^\circ$ with automated computer acquisition (Fig. 4). Each specimen was scanned along two orthogonal directions. Instrumental broadening was corrected by measuring a reference sample of macrocrystalline aluminum. The surfaces of the specimens for X-ray analysis were prepared exclusively by mechanical treatment.

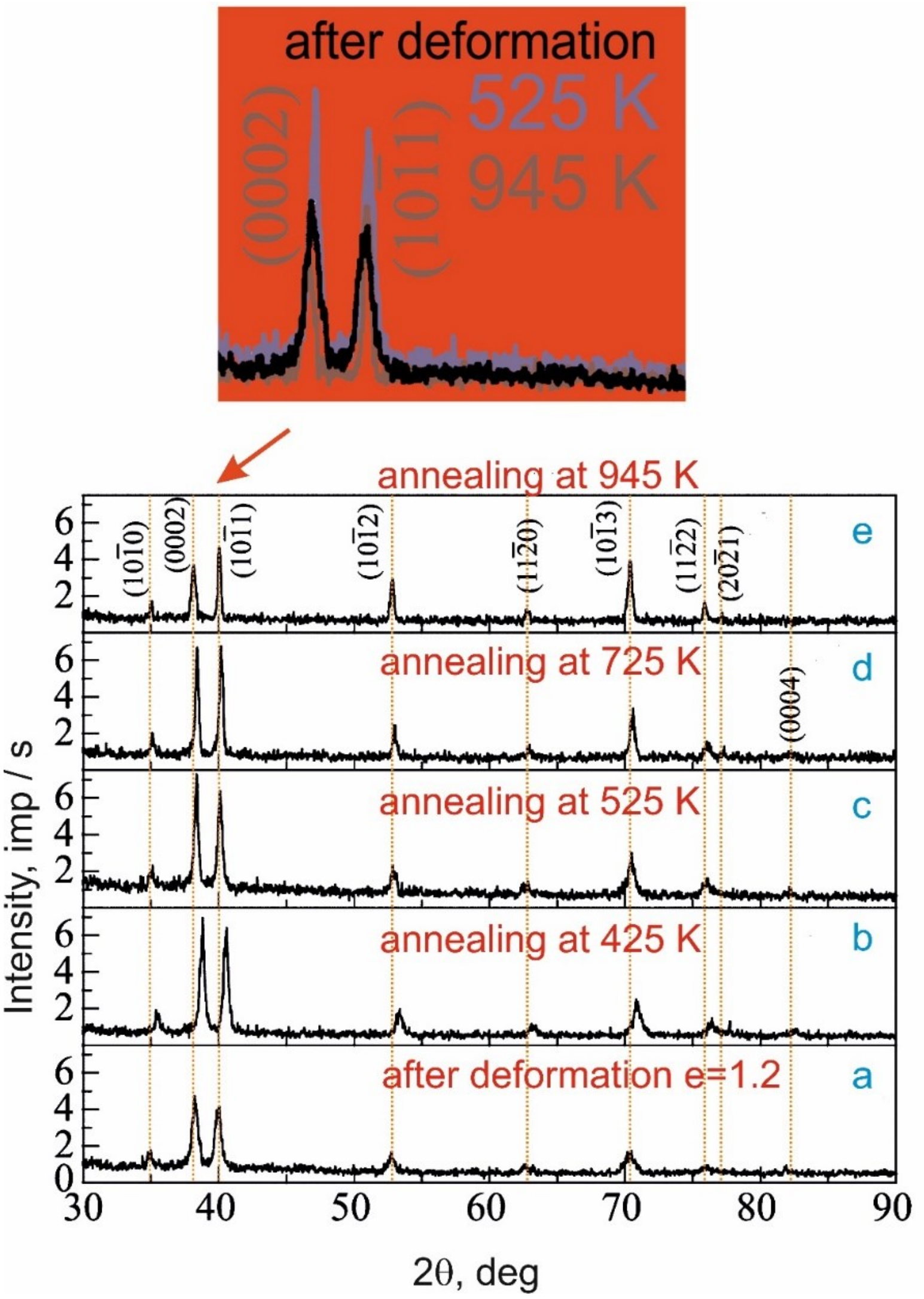

**Fig. 4.** Diffraction patterns of VT1-0 titanium recorded immediately after deformation (a) – state **1**; and following annealing at 425 K (b) – state **2**; 525 K (c) – state **3**; 725 K (d) – state **4**; and 945 K (e) – state **5**. The inset illustrates a typical transformation of the diffraction profiles, expressed as the broadening of diffraction maxima.

## 3. Results and Discussion

Structural parameters were evaluated through analysis of diffraction peak intensity, profile shape, and peak width [23]. The X-ray coherent scattering region size *L* within crystallites, together with the degree of microstrain $\Delta\varepsilon$, was determined using an approximation approach based on the broadening of diffraction maxima [24-29]. The experimental breadth $\beta_{\mathrm{exp}}$ of the Bragg peaks reflects a combination of instrumental and sample-related contributions [23]. To eliminate these distortions, instrumental broadening was quantified by constructing a diffraction pattern of a standard reference material. The corrected physical value $\beta_{hkl}$ of the Full Width at Half Maximum (FWHM) for the alloy under study was then obtained using the following relation:

$$\beta_{hkl} = \beta_{\mathrm{exp}} - \beta_{\mathrm{instr}}, \tag{1}$$

here, $\beta_{\mathrm{exp}}$ denotes the experimental FWHM of the XRD peak; while $\beta_{\mathrm{instr}}$ represents the instrumental contribution to the FWHM, determined using standard reference materials for crystallite size evaluation.

It is well established that peak broadening $\beta_{hkl}$ is influenced not only by finite particle size but also by mechanical stresses arising from defects in the crystal lattice. According to the Williamson-Hall (W-H) model [24, 25], the observed broadening $\beta_{hkl}$ is expressed as the additive contribution of crystallite size effects $\beta_L$ and the lattice microstrain $\beta_\varepsilon$:

$$\beta_{hkl} = \beta_L + \beta_\varepsilon. \tag{2}$$

As is widely recognized, the broadening contribution associated with crystallite size $\beta_L$ is described by the Debye–Scherrer equation [26]:

$$\beta_L = \frac{K\lambda}{\langle L\rangle \cdot \cos\theta}, \tag{3}$$

here, $\lambda$ denotes the X-ray wavelength; *K* is a dimensionless shape factor, typically close to unity [26], $\langle L\rangle$ represents the average size of the ordered crystalline domains, which may be equal to or smaller than the grain size, and in turn smaller than the overall particle size (crystallite size); and $\theta$ is the Bragg angle.

Crystal imperfections and strain-induced peak broadening within the framework of the uniform deformation model are related through the following expression $\langle\varepsilon\rangle \propto \dfrac{\beta_\varepsilon}{\tan\theta}$:

$$\beta_\varepsilon = 4\langle\varepsilon\rangle \tan\theta = 4\langle\varepsilon\rangle \frac{\sin\theta}{\cos\theta}, \tag{4}$$

here, $\langle \varepsilon \rangle$ denotes the mean square lattice micro strain. Thus

$$\beta_{hkl} = \beta_L + \beta_\varepsilon = \frac{K\lambda}{\langle L \rangle \cdot \cos\theta} + 4\langle \varepsilon \rangle \tan\theta \quad \Rightarrow \quad \frac{\beta_{hkl} \cdot \cos\theta}{\lambda} = \frac{K}{\langle L \rangle} + \frac{4\langle \varepsilon \rangle \cdot \sin\theta}{\lambda}. \tag{5}$$

An important feature of Eq. (5) is its dependence on the diffraction angle $\theta$. Debye-Scherrer equation follows only a $\frac{1}{\cos\theta}$ dependency but not $\tan\theta$ as W-H model. Consequently, both crystallite size and strain can be estimated by plotting $\frac{\beta_{hkl} \cdot \cos\theta}{\lambda}$ against $\frac{4 \cdot \sin\theta}{\lambda}$ ($\beta_{hkl}$ in radian and $\theta$ in degree) for all diffraction peaks and fitting a straight line. The intercept of this line corresponds to the crystallite size contribution $\frac{K}{\langle L \rangle}$, whereas the slope represents the microstrain $\langle \varepsilon \rangle$. The W-H analysis leads to the empirical estimates of the average value of the ordered crystalline domains (particle size or coherent scattering regions) - Fig. 5a; and average value of the lattice micro strain - Fig. 6.

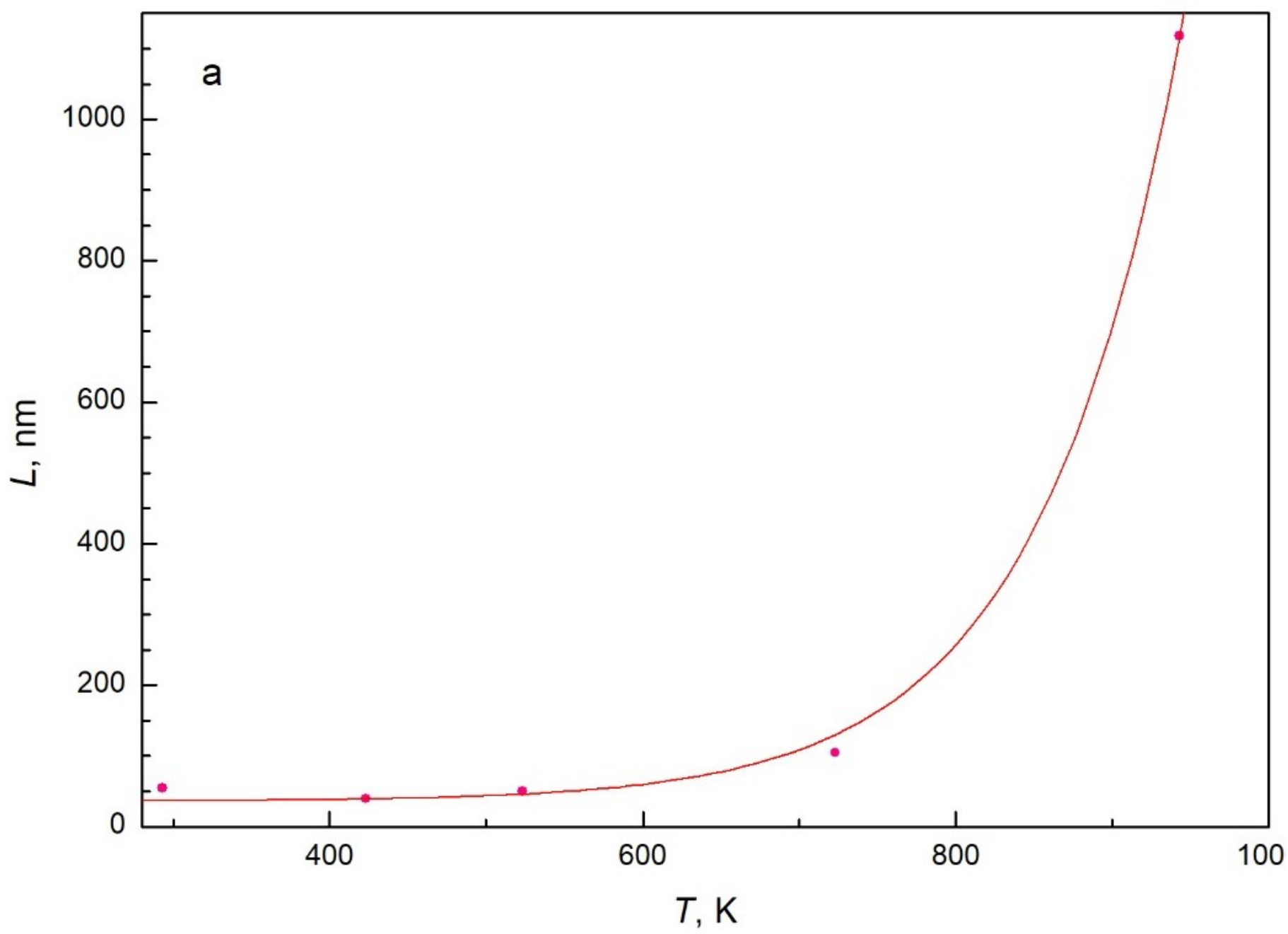

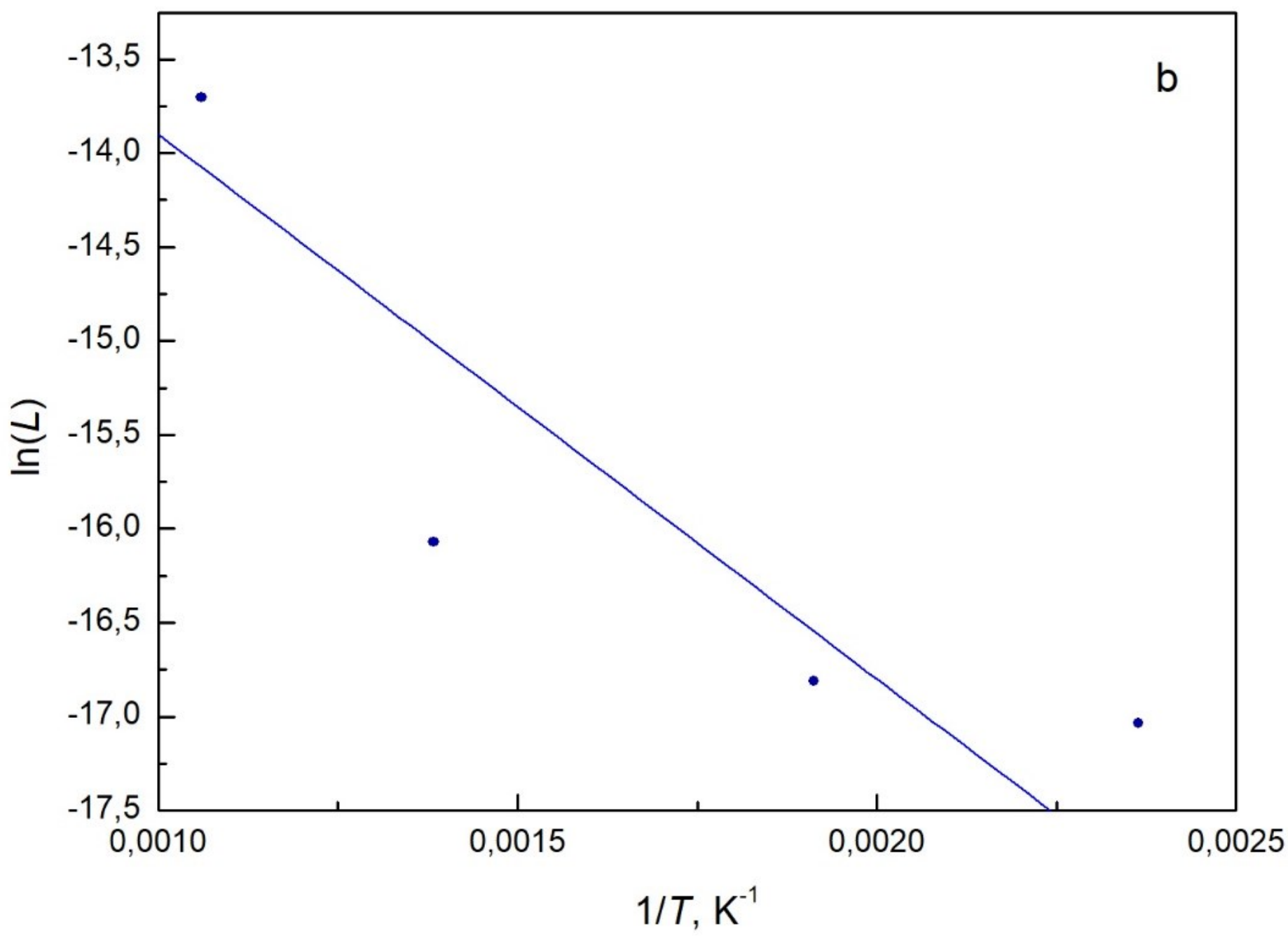

**Fig. 5.** The evolution of the characteristic grain size (a) and kinetic characteristics of grain growth (b). The solid line in Fig. 5 a represents the analytical approximation (9), while the solid line in Fig. 5b corresponds to the theoretical dependence (14).

Both of parameters ($\langle L\rangle$ and $\langle\varepsilon\rangle$), obtained from W-H analysis, can be employed to estimate the dislocation density [25]. The dislocation density $\rho_L$ derived from the particle size will be determined according to the following ratio:

$$\rho_L = \frac{3m}{\langle L\rangle^2}, \tag{6}$$

here, $m$ denotes the number of dislocations per block face, the value of $m$ must either be determined experimentally or assumed. In our calculations, we adopt $m = 1$; this value corresponds to the minimum density of edge dislocations and is applicable for a structural state in which the dislocations are distributed in an almost random manner [26].

The calculation of $\rho_\varepsilon$ for the hexagonal lattice was performed using the method proposed in [30]:

$$\rho_\varepsilon = \frac{9\cdot\pi}{16}\cdot\frac{\langle\varepsilon\rangle^2}{b^2}. \tag{7}$$

The overall average dislocation density $\langle\rho\rangle$ (Fig. 7), was evaluated as the geometric mean [31-35]:

$$\langle\rho\rangle = \sqrt{\rho_L\cdot\rho_\varepsilon}. \tag{8}$$

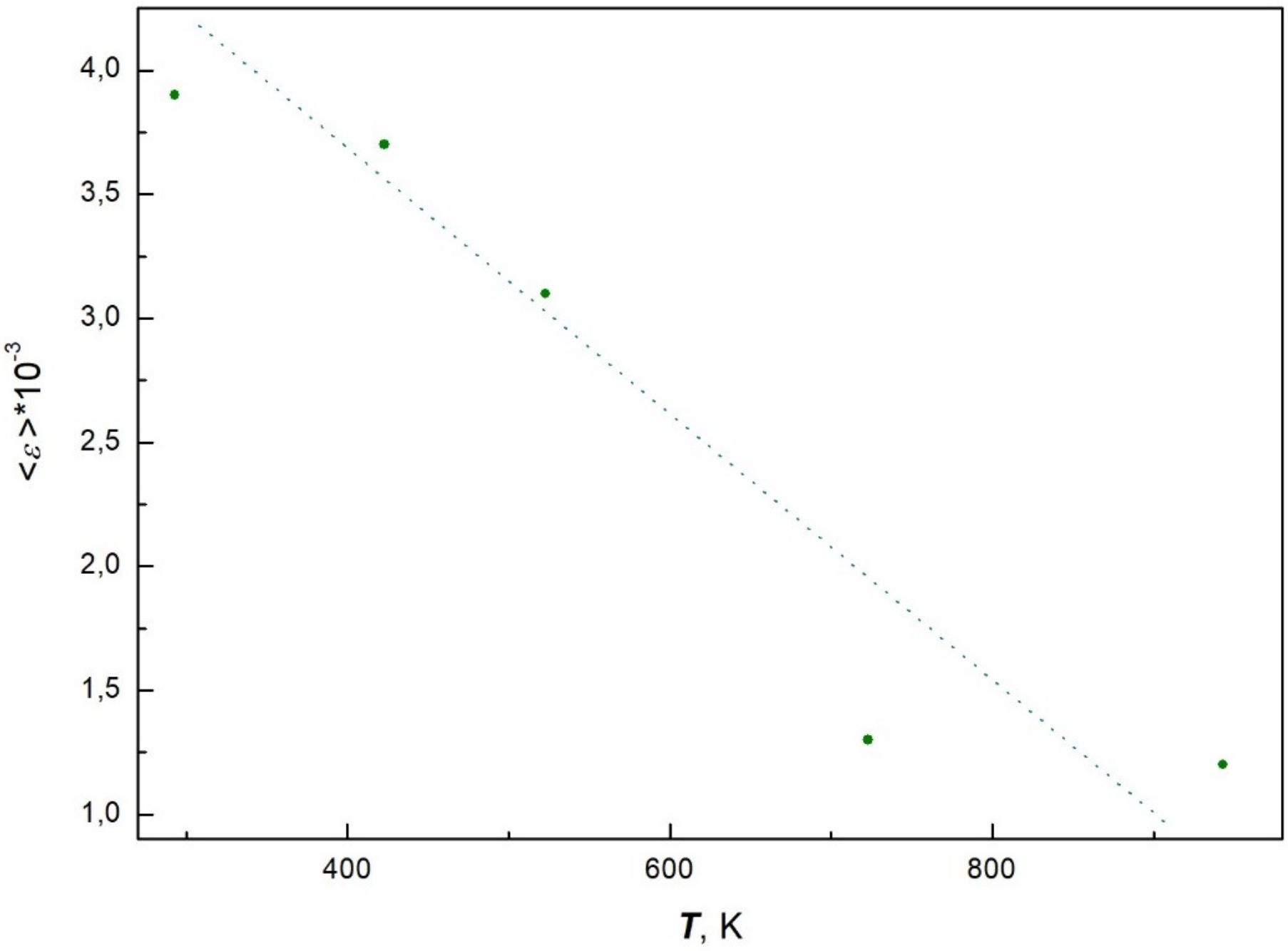

**Fig. 6.** The evolution of the average microstrain $\langle \varepsilon \rangle$ during isothermal annealing at 425 K, 525 K, 725 K, and 945 K.

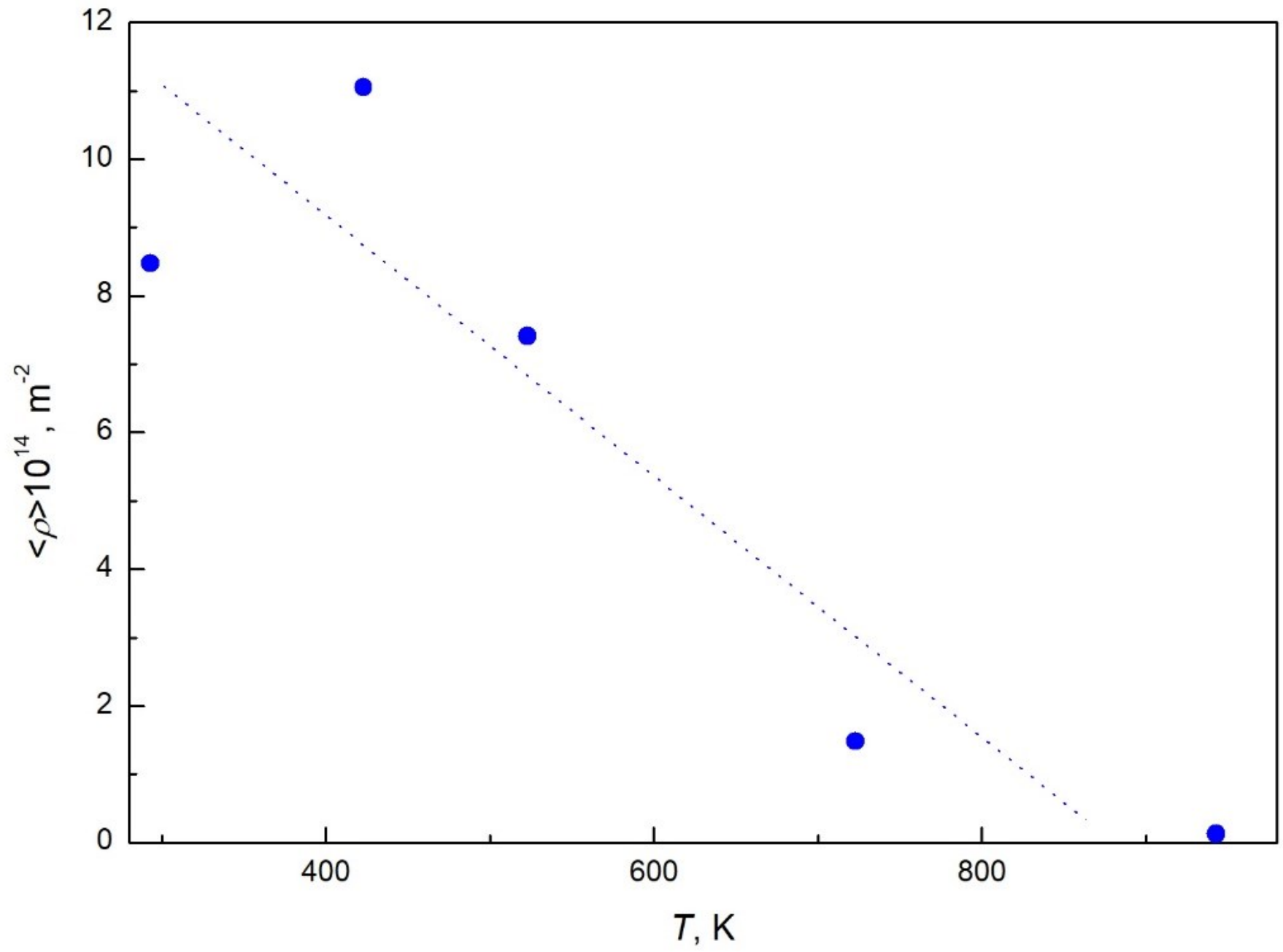

**Fig. 7.** The evolution of the average dislocation density $\langle \rho \rangle$ during isothermal annealing at 425 K, 525 K, 725 K, and 945 K.

Table 1 presents the results of X-ray diffraction pattern analysis for initial state, immediately after cryodeformation at $|e| = 1.2$.

**Table 1.**

Structural parameters of the VT1-0 titanium specimen after CTR derived from X-ray diffraction analysis. The size $L_0$ of regions exhibiting X-ray coherent scattering in crystallites (Coherent Scattering Regions - CSRs), as well as the level of microdeformation $\Delta\varepsilon$ and the unit cell parameters ($a$, $c$) were taken from [14, 36].

| parameter | | structural state | | | | |
|---|---|---|---|---|---|---|
| | | state **1** | state **2** | state **3** | state **4** | state **5** |
| cell parameters | $a$, Å | 2.967 | | | | |
| | $c$, Å | 4.709 | | | | |
| $L_0$, nm | | 55 | 40 | 50 | 105 | 1118- |
| $\Delta\varepsilon$ $10^{-3}$ | | 3.9 | 3.7 | 3.1 | 1.3 | 1.2 |
| $\langle\rho\rangle$ $10^{14}$, m$^{-2}$ | | 8.47 | 11.05 | 7.41 | 1.48 | 0.13 |

The analysis shows that annealing promotes an increase in the average size of CSR determined from X-ray diffraction (Fig. 5a). The growth of CSR becomes evident with rising annealing temperature, particularly after treatment at 725 K. At 525 K, annealing produces only a minor enlargement of the crystallite size, reaching about 50 nm, while simultaneously relieving most of the internal stresses introduced during cryorolling. Further annealing at 725 K results in a twofold increase in grain size, up to ~ 100 nm. Finally, annealing at 945 K leads to complete recrystallization, thereby eliminating the nanostructured state of the alloy.

Annealing thus relieves internal stresses and promotes grain (crystallite) growth; however, it has been established that it does not alter the lattice parameters, meaning no phase transition is observed. Consequently, the lattice parameter remains unchanged with annealing. A more detailed analysis of lattice parameters was not carried out in this study.

In contrast to coarse-grained samples, where internal stresses are determined both by grain size and by the intragranular substructure, in nanocrystalline titanium the magnitude of internal stresses is governed exclusively by grain size. As shown in the works of [14 and 37], nanosized titanium grains possess a nearly perfect internal structure, which indirectly supports the assumption that dislocation accumulation - responsible for lattice distortions - is impossible. Low-temperature annealing (325–425 K) leads to an increase in microhardness and strength characteristics of nanostructured titanium [19, 38, 39]. This distinctive behavior of nanostructured metals is attributed to the rearrangement of grain boundaries (without significant grain growth), which results in observable strengthening and a corresponding change in dislocation density.

The experimental data points in Fig. 5a can be accurately approximated by an analytical relationship:

$$L = L_0 + A_1 \cdot \exp\left(\frac{T}{A_2}\right), \tag{9}$$

here, $L_0$ denotes the initial CTR size (Table 1) , while $A_1 = 0.03143$ nm and $A_2 = 90.285$ K$^{-1}$ are fitting parameters. The exponential nature of this relationship indicates that the kinetics of CSR growth can be described within the classical Lifshitz – Slyozov - Wagner (LSW) framework, which accounts for grain growth rates and the diffusion-controlled motion of atoms during grain boundary migration [40]. Thus, the rate of change of the average grain size $L$ with time $t$ is proportional to the driving force of the process, namely the curvature of the grain boundaries $1/L$ :

$$\frac{dL}{dt} = \frac{K}{L^{n-1}}, \tag{10}$$

here, $K$ is the proportionality constant, which is strongly temperature-dependent, and $n$ is the grain growth parameter (commonly taken as $n = 2$ for pure metals, or $n = 3$–4 for alloys and nanomaterials). By integrating this equation over time from 0 to $t$, and assuming that the initial nanoscale grain size $L_0$ is negligible compared to the grown grain size $L$, we obtain the classical parabolic law describing the kinetics of isothermal grain growth [41]:

$$L^n - L_0^n = k \cdot t \Rightarrow L^n \approx k \cdot t \Rightarrow L = (k \cdot t)^{1/n}. \tag{11}$$

Since grain boundary migration is a diffusion-controlled process involving discrete atomic jumps, the rate constant $k$ follows the Arrhenius law:

$$k = k_0 \cdot \exp\left(-\frac{Q_r}{R \cdot T}\right), \tag{12}$$

here $R = 8.314 \frac{\text{J}}{\text{mol}\times\text{K}}$ denotes the universal gas constant, $T$ is the annealing temperature, and $k_0$ represents the pre-exponential factor, which reflects the maximum possible rate of the process at “infinite” temperature, when thermal energy is sufficient to overcome any barrier. In the case of grain boundary migration $k_0$, depends on the atomic structure of the material and incorporates several parameters: the attempt frequency of an atom crossing the grain boundary (on the order of the Debye frequency ≈ $10^{13}$ $s^{-1}$); the displacement distance of the boundary during a single atomic jump (comparable to the lattice parameter, b ≈0.3 nm); and the activation entropy $\Delta S$, which characterizes the change in structural disorder at the moment of atomic transition across the boundary ( $k_0 \propto \exp\left(\frac{\Delta S}{R}\right)$).

Based on equations (11) and (12), and under the assumption of a fixed duration of isothermal annealing, we obtain:

$$L = A_3 \cdot \exp\left(-\frac{Q_r}{nRT}\right), \tag{13}$$

here $A_3 = \left(k_0 \cdot t\right)^{1/n}$ is pre-exponential factor (constant), and $Q_r$ - the apparent activation energy of grain growth (J/mol). Thus, by replotting the experimental data from Fig. 4 in semi-logarithmic coordinates:

$$\ln\left(L\right) = \ln\left(A_3\right) - \frac{Q_r}{n \cdot R} \cdot \frac{1}{T}, \tag{14}$$

it becomes possible to evaluate the kinetic characteristics of grain growth and the evolution of the characteristic grain size (Fig. 5b).

Cryogenic deformation generates an extremely high dislocation density and induces strong local stresses. Grain boundaries in such material are in a non-equilibrium state, while the volume fraction of triple junctions is remarkably large. These triple junctions act as effective barriers to grain boundary migration, so grain growth is limited by the mobility of the junctions themselves and by boundary diffusion under defect-impeded conditions. In this case, the growth exponent $n$ can be taken as ≈ 4. This leads to an estimated activation energy $Q_r \simeq$ [illegible] (Fig. 5b), which corresponds to typical values for grain boundary diffusion in α-Ti [42].

Within the LSW framework, one of the key conclusions is that the most rapid changes in CSR size and microstrain occur at the early stages of annealing (at temperatures below 425 K, which were not investigated in this work). At the higher temperatures studied here, the growth of these parameters is considerably slower, and the exponential dependence tends to approach a quasi-linear form - exactly as observed experimentally in Figs. 6 and 7. In other words, at lower annealing temperatures grain growth itself is not significant, yet the rate of structural rearrangement is high: grain boundaries undergo active reorganization and microstrains are relieved. These processes demand less energy than diffusion-controlled grain growth, which explains their intensity even under relatively mild heating. Consequently, low-temperature annealing produces the most vigorous structural relaxation, driven by the system's tendency to eliminate excess defects and internal stresses. At elevated temperatures, once the defect density has diminished, the kinetics slow down and the mechanism shifts toward conventional diffusion-mediated grain growth.

In [15] it is shown that X-ray investigations reveal the presence of an X-ray amorphous phase; however, crystallites smaller than 15 nm cannot be distinguished from an amorphous phase by X-ray diffraction, whereas TEM makes it possible to differentiate a true amorphous phase and directly observe crystallites of any size. It should be noted, however, that TEM provides local information on the morphology and distribution of crystallites, while X-ray diffraction delivers integral data over large sample volumes, including the evaluation of microstresses, texture, and lattice parameters.

## 4. Conclusions

1. It has been shown that plastic deformation through rolling results in substantial grain subdivision in the original titanium material. In specimens deformed at 100 K, the substructure is dominated by Coherent Scattering Regions (CSRs) with dimensions of about 30-50 nm. Cryogenic rolling facilitates the emergence of nanoscale domains owing to the activation of multiple twinning systems.

2. The conducted analysis demonstrates that annealing significantly influences the structural relaxation of cryorolled commercially pure titanium. At moderate temperatures (525 K), annealing produces only a slight increase in crystallite size (~50 nm) while effectively relieving internal stresses. A more pronounced grain growth occurs at 725 K, where the average size nearly doubles to ~100 nm, consistent with diffusion-controlled boundary migration described by the Lifshitz – Slyozov - Wagner framework. Complete recrystallization is achieved at 945 K, eliminating the nanostructured state.

3. The apparent grain growth activation energy of the nanostructured cryodeformed titanium within the temperature range of 425-725 K was found to be 97 kJ/mol. This value strictly corresponds to the lower bound of the reported activation energy for grain boundary diffusion in α-Ti (97-130 kJ/mol), demonstrating that the grain growth kinetics in this regime are predominantly governed by grain boundary atom migration.

## Acknowledgments

The authors gratefully acknowledge the support of the National Academy of Sciences of Ukraine, the Ministry of Education and Science of Ukraine, and the Polish Academy of Sciences.

This paper is dedicated to the memory of Professor Vasilij Natsik, whose profound intellectual contributions and unwavering commitment were vital to the completion of this research. Professor Natsik passed away during the preparation of this manuscript. Through this work, we honor his lasting legacy and invaluable impact on the scientific community.

## Author Contributions

Conceptualization, Y.S.; Methodology, Y.S., N.G., D.H. V.Z.; Validation, Y.S.; Formal Analysis, V.Z. and P.Z.; Investigation, Y.S., N.G., D.H., T.B, N.A., V.M., R.S., A.S., Y.P.; Resources, N.G. and D.H.; Data Curation, Y.S.; Writing - Original Draft Preparation, Y.S.; Writing - Review & Editing, Y.S.; Visualization, V.Z. and Y.S.; Supervision, Y.S.; Project Administration, Y.S.; Funding Acquisition, Y.S.

## Data Availability Statement

The data presented in this study are available upon request from the corresponding author.

## Disclosure statement

No potential conflict of interest was reported by the authors.

## Funding

This work was partly supported by the Projects No. 0122U001504 and No. 0124U000272; internship within the framework of scientific cooperation between the National Academy of Sciences of Ukraine and the Polish Academy of Sciences.

## Employment

B.Verkin Institute for Low Temperature Physics and Engineering, National Academy of Sciences of Ukraine, Institute of Low Temperature and Structure Research, Polish Academy of Sciences.

## ORCID

Yuri Semerenko https://orcid.org/0000-0003-2268-0572

Nikolai Galtsov https://orcid.org/0000-0002-6240-4006

Diana Hurova https://orcid.org/0000-0002-8054-4758

Tamara Bednarchuk https://orcid.org/0000-0001-8741-6057

Viktor Zoryansky https://orcid.org/0000-0002-7473-3288

Peter Zinoviev https://orcid.org/0000-0002-9195-5018

Vladyslav Moskalenko https://orcid.org/0000-0003-0920-7609

Ruslan Smolianets https://orcid.org/0000-0002-0031-9332

Yuliya Pohribnaya https://orcid.org/0000-0002-1529-7260

Nataly Aksenova https://orcid.org/0000-0001-8194-836X